\documentstyle[referee,epsfig]{mn}
\input epsf

\title[Bulge]{The morphology and luminosity function
of the Galactic bulge from TMGS star counts}

\author[L\'opez-Corredoira et al.]{M. L\'opez--Corredoira$^1$, 
F. Garz\'on, P. Hammersley, T. Mahoney, X. Calbet
\\
Instituto de Astrof\'\i sica de Canarias, E-38200  La 
Laguna, Tenerife, Spain
\\
$^1$ e-mail: martinlc@iac.es}

\date{Accepted xxxx.
      Received xxxx;
      in original form xxxx}

\begin{document}

\maketitle

\begin{abstract}
The bulge of the  Galaxy is analysed by inverting  $K$-band star 
counts from the Two-Micron Galactic Survey  in a number of off-plane  regions. A total area of about $75$ square degrees of sky is analysed. Assuming a non-variable luminosity function within the bulge,
we derive the top end of the $K$-band luminosity function and 
the stellar  density function, whose morphology is fitted to  triaxial ellipsoids.

The luminosity function shows a sharp decrease brighter than  $M_K=-8.0$ when compared  with the disc population. By fitting  ellipsoids, we find that
the bulge is triaxial with the major axis in the plane at an angle with
line of sight to the Galactic centre of $12^\circ$ in the first quadrant. The axial ratios 
are $1:0.54:0.33$ and the distance of the Sun from the centre of the triaxial ellipsoid is $7860$ pc.

The best fit for the stellar density, assuming an ellipsoidal distribution, 
is $D(t)=1.17(t/2180)^{-1.8}\exp(-
(t/2180)^{1.8})\ {\rm stars\ \ pc^{-3}}$, for $1300<t<3000$,
where $t$ is the distance along the major axis of the  ellipsoid in parsecs.
\end{abstract}


\section{Introduction}

Among the many aspects of the  bulge of the Galaxy that are still unknown,
mainly because of the high extinction due to interstellar gas and dust,
is the near-infrared luminosity function, knowledge of which is provided  only by observations in Baade's window (Frogel \& Whitford 1987; Davidge 1991; 
De Poy et al. 1993; Ruelas-Mayorga \& Noriega-Mendoza 1995; Tiede, Frogel \&
Terndrup 1995). In the visible, Gould (1997) uses the {\it Hubble Space Telescope} $V$ and $I$ bands. However, from extrapolation from Baade's
 and other clear windows to the whole bulge may not be appropriate,
in particular because these are `special' regions.
Furthermore, these regions are very small, containing relatively few stars, and hence give very poor statistics at the brighter magnitudes. The bright end of the luminosity
function is very important in order to determine the age of the population.

Many authors find non-axisymmetry in the Galactic bulge 
(Weiland et al. 1994; Feast \& Whitelock 1990 ; etc.) with  a negligible out-of-plane tilt (Weiland et al. 1994) and
giving more counts in the positive
galactic longitudes than in the negative ones. However, other authors 
(Ibata \& Gilmore 1995) claim that axisymmetry is suitable. 

This paper examines the $K$ distribution of bulge sources in the Two-Micron
Galactic Survey (TMGS) 
though the inversion of the fundamental equation of stellar statistics
using Lucy's algorithm (Lucy 1974).

\section{Star counts and inversion procedure}

The $K$-band star counts are from the  TMGS (Garz\'on et al. 1993), which has covered about $350$ square degrees of sky and detected some 
700000 sources in or near the Galactic plane. The survey has a completeness
limit in the bulge region of $m_K\approx +9.0$, except for the regions near the Galactic centre
where confusion reduced this by about half a magnitude.

For this study we use 71 regions in the Galactic bulge ($10^\circ>|b|>2^\circ$, $|l|<15^\circ$) taken coming from strips with $\delta =-30^\circ$, $\delta =-22^\circ$ 
and $\delta =-16^\circ$. The TMGS strips are at constant declination, which means that they cross
the Galactic plane at a significant angle; hence the three strips sample a wide
range of lines of sight through the bulge. The total sky coverage is some 
$75$ deg$^2$ of sky. The 
area near the Galactic plane was not used in order to avoid components 
which belong neither to the bulge nor to the disc (e.g. spiral arms)
and the high and variable extinction close to the plane.  The
outer limits were set so that the bulge-to-disc stellar ratio was still acceptable, i.e. there are enough bulge stars in comparison with disc stars.

In the areas considered, the contribution to the star counts will be 
primarily from the disc and bulge. In order to isolate the bulge component 
a model disc was subtracted from the total counts. The model developed
by us was based on Wainscoat et al. (1992), which has been used because
it provides a 
good fit to the TMGS counts in the region where the disc dominates (Cohen 1994b;
Cohen et al., in preparation). The revision to Wainscoat
et al. (1992) given in Cohen (1994a) does not significantly alter
the form of the disc in the area of interest.
It was also expected that the model 
would give an adequate  fit for the bulge counts; this, 
however, was not the case.

For each of the 71 regions, centred on Galactic coordinates $(l,b)_i$, where $i$ is the field number, the cumulative stellar star counts towards the bulge, $N_K$, expressed in rad$^{-2}$, follows

\[
N_{K,{\rm bulge}}(m_K)=N_K(m_K)-N_{K,{\rm disc}}(m_K)=
\]\begin{equation}
\int_0^\infty \Phi_{K,{\rm bulge}} (m_K+5-5\log _{10} r -a_K(r))
D_{\, \rm bulge}(r) r^2 dr
\label{sc_acum}
,\end{equation}
where
$\Phi _{K,{\rm bulge}}(M)=\int _{-\infty}^{M}\phi _{K,{\rm bulge}}(M)dM$,
$\phi $ is the normalised luminosity function ($\int _{-\infty}^\infty$
$\phi (M)dM=1$),
$D$ is the density,
and $a_K$ is the extinction in the line of sight. For the extinction we
have followed Wainscoat et al. (1992), who assume that the extinction 
has an exponential distribution with the same scale length as the old disc
and a scale height of 100 pc. This is normalised to give  
$A_K=\frac{da_K}{dr}=0.07$ mag kpc$^{-1}$ in the solar neighbourhood. As the areas of interests are off the plane  the extinction to the bulge sources is between  $0.05$ to 
$0.5$ mag at $K$. Garz\'on et al. (1993)  showed that, away from the
plane, the $K$ extinction varies smoothly and has 
little effect on the distribution
of the star counts. In fact, the TMGS histograms
in Garz\'on et al. (1993) off the plane show that the extinction cannot
be patchy.

With the change of variables
$\rho _K=10^{0.2a_K(r)}r$
and
$\Delta _K=D(r)\frac{r^2dr}{\rho _K^2 d\rho _K}$
we transform the equation (\ref{sc_acum})
of counts in the bulge into

\begin{equation}
N_K(m_K)=\int_0^\infty \Phi_K (m_K+5-5\log _{10} \rho _K )
\Delta_K(\rho _K) \rho _K^2 d\rho _K
\label{sc_acum_fic}
.\end{equation}

The density is obtained by inverting  this equation:
$\Delta $ is the unknown function and $\Phi $ is the kernel of
a Fredholm integral equation of the first kind (see Trumpler \& Weaver 1953, p. 96).

When  the luminosity function
$\Phi $, is the unknown instead of $\Delta $, then we can make a new change of variable
$M_K=m_K+5-5\log _{10}\rho _K$ and we obtain a new first kind of Fredholm
equation:

\[
N_{K}(m_K)=200({\rm ln}\ 10)10^{\frac{3m_K}{5}}
\times \]\begin{equation}\times
\int_{-\infty }^\infty \Delta_K(
10^{\frac{5+m_K-M_K}{5}}) 10^{\frac{-3M_K}{5}}
\Phi_K (M_K)dM_K
\label{lumin_inc}
,\end{equation}
where $\Phi $ is now the unknown function and $\Delta _K$ is the kernel.

Both integral equations are inverted using
Lucy's statistical method (Lucy 1974). This method is fairly 
insensitive  to the high-frequency fluctuations and in our tests 
with known functions, which are similar to that of the bulge, gave good results
(note: this method would not be applicable to the disc as a whole).

The above equation can be solved for either the luminosity function or the
density function, but not both simultaneously. A simple comparison of the Wainscoat et al. (1992)
model with the TMGS counts suggested that while there were problems with the
bulge luminosity function, the density function gave an adequate starting value for the iteration. Therefore it was decided to solve first for the average 
luminosity function using the Wainscoat et al. (1992) bulge density. 

We have made  the assumption that  the luminosity function  is independent
of the position. This assumption is suspect (see Frogel 1988, section 3) since the observed metallicity gradient might affect the luminosity of an AGB star, although not the non-variable M-giants
whose bolometric luminosity function is nearly independent of the
latitude (Frogel et al. 1990). Some authors claim that there is a 
population gradient (Houdashelt 1996), whilst others do not (Ibata \& Gilmore 1995, who argue that there is no detectable abundance gradient in the Galactic bulge over the galactocentric range from 500 to 3500 pc).
While our assumption may not be not strictly true,  it is still a reasonable
approximation. We, therefore, assume that the variation of the bulge luminosity
function in between about $250$ pc and $1200$ pc from the galactic plane 
is small.

With this averaged luminosity function, we inverted (\ref{sc_acum_fic}) to derive a new density distribution. In this step we used the  37 regions with the highest counts as the determination of the density is  
more sensitive to noise.
The inversion of the luminosity function is more stable because
the density distribution is sharply peaked and so the kernel in
(\ref{lumin_inc}) behaves almost as a Dirac delta function: the shape of the
density distribution does not significantly affect the shape
of the luminosity function. The new density was then used to
improve the luminosity function, etc. The whole process was iterated three 
times which was enough for the results to stabilize as can be seen
in the Fig. \ref{Fig:luminosity}: we see how the result of the third iteration
is very close to the first; i.e. stabilization is reached
in the first iterations.

The functions of interest are
$\phi $, the derivative of $\Phi $,
and $D$, related to $\Delta $ by the change of variable expressed above.

\section{The Top end of the $K$ luminosity function}

\begin{figure}
\begin{center}
\mbox{\epsfig{file=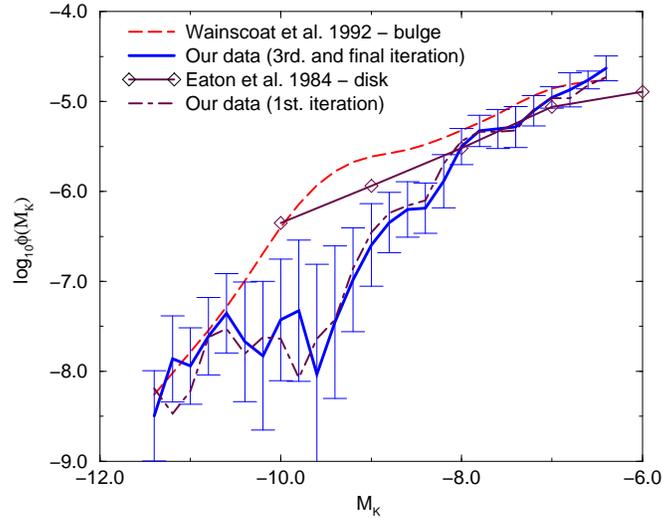,height=8cm}}
\end{center}
\caption{Luminosity function of the bulge stars in the $K$ band: the 
thick solid line is the third and final iteration.
Comparisons with the same luminosity function in the first iteration,
Wainscoat et al. (1992) in the bulge and Eaton, Adams \& Gilels (1984)
in the disc are also provided.}
\label{Fig:luminosity}
\end{figure}

After three iterations the luminosity function was independent of the position  $(l,b)_i$  and  stable.
The obtained luminosity function is shown in the Fig. \ref{Fig:luminosity}. 
The coincidence of our luminosity function with that of
Wainscoat et al. (1992) for the faintest stars is due to the fact
that we used their luminosity function to initiating the
iteration process. Practically speaking, this overlapping corresponds
to an effective normalization to the Wainscoat et al. (1992) luminosity
function in the range $M_K>-7$.

 Fig. \ref{Fig:luminosity} shows that  for  $-10< M_K <-7$ 
the bulge luminosity function
is significantly lower  than that of the disc (Eaton, Adams \& Gilels 1984).
Hence, the density  of very bright stars 
in the bulge is much less than in the disc. Fainter than $M_K=-7$ the 
luminosity functions of the disc and the bulge coincide,
in agreement with Gould (1997). The luminosity function for  $-10<M_K<-7$ is significantly 
below  the synthesized luminosity function assumed by Wainscoat et al. (1992)
for the bulge in their model of the Galaxy.
The discrepancy could arise from their not having 
taken into account that the brightest stars in the bulge are up to 2
magnitudes fainter than the disc giants (Frogel \& Whitford 1987).
This would shift the luminosity function to the right.

Comparison with bolometric luminosity function obtained by other
authors (see references in the introduction)
is not possible since we do not have the bolometric corrections. Also,
in most of cases the magnitude interval  is different.
Tiede, Frogel \& Terndrup (1995) provide, by combining data from
different articles, the luminosity function in the $K$-band as a function of the
apparent magnitude in the range $5.5<m_K<16.5$. The brightest
magnitudes are taken from Frogel \& Whitford (1987). The comparison with
our luminosity function is not direct since they have not normalized
their luminosity function to unity; moreover, they have not taken
into account the narrow but non-negligible
dispersion of distances. In Fig. 16 of Tiede, Frogel \& Terndrup (1995) 
there is a 
fall-off of the luminosity function for $m_K\le 6.5$ or in Fig. 
18 of Frogel \& Whitford (1987) for $M_{bol} \le -4.2$,
which could be comparable with that
of our luminosity function at $M_K \approx -8.0$. However, because of the far
larger area covered by the TMGS, the error for the brightest magnitudes is 
far lower in this paper, the result being pushed well above the noise; this is not the case for Frogel \& Whitford (1987).

The presented luminosity function for very bright stars 
(lower than $M_K \sim -9.5$)
is of low accuracy. The number of bulge stars in this range 
is very small, small errors
due to contamination from the spiral arms
will mean that the luminosity function is overestimated.

The bulge is older than the disc. However,
the comparison with the halo globular
clusters ages remains open.
These data could help to investigate 
the age of the bulge by 
comparison with theoretical models of stellar evolution.

\section{Bulge morphology}

 The morphology of the bulge can be examined by fitting the isodensity 
surfaces to  $D(\vec{r}) =D(r,l,b)$. 
We fitted three-dimentional ellipsoids to  $20$ isodensity surfaces
(from $0.1$ to $2.0$ star pc$^{-3}$, in steps of $0.1$)
with four free parameters: $R_0$, the Sun-Galactic centre distance (the
ellipsoids are then centred on this position);
$K_z$ and $K_y$, the axis ratios with respect the major axis $x$;
and $\alpha $ the angle between the major axis of the triaxial
bulge and the line of sight to the Galactic centre
($\alpha $ between $0^\circ $ and $90^\circ $ is where the tip of the major
axis lies in the first quadrant).
These ellipsoids have two axes in the Galactic plane,
($x$ and $y$) and the $z$ axis which is perpendicular, we have ignored
a possible tilt out of the plane.

The four averaged  parameters have been fitted for the 20
ellipsoids and the results are:

$R_0=7860 \pm 90$ pc 

$K_z=3.0 \pm 0.9$ 

$K_y=1.87 \pm 0.18$ 

$\alpha =12 \pm 6$ deg.

We can also express  the axial ratios as $1: 0.54: 0.33$.
The errors are calculated from the average of the ellipsoids,  and so do
not include possible systematic errors (for example: subtraction of the disc,
contamination from other components, procedure of the inversion,...).
Hence, the true errors are larger that stated but tests suggest
that they do not alter the general findings of this paper.
These numbers indicate that  the bulge is triaxial with the major axis
 close to the line of sight towards the Galactic centre.
The error in $K_z$ is quite large and  is due to a
non-constant axial ratio of the ellipsoids.  There is a trend towards 
increasing $K_z$ with  proximity to the centre, i.e. the outer bulge is more circular
 than the inner bulge.

In general the result presented here are in agreement with those from
other authors. The projection of an ellipsoid of the above 
characteristics on to the sky,
as viewed from the position of the Sun, gives an ellipse with axial ratio
$1.7 \pm 0.5$ (i.e. $1:0.58$). This is compatible with the value of $1:0.6$ obtained by
Weiland et al. (1994).
Dwek et al. (1995) give higher  eccentricity values for the  axial ratios
($1:0.33:0.22$), but the angle $\alpha =20 \pm 10$ deg is compatible with the  
value given here. From the dynamic model, assuming a 
gas ring in steady state, Vietri (1986) finds axial ratios of $1: 0.7: 0.4$, 
which are close to our result. Binney et al. (1991) find  $\alpha =16$ deg. for a bar, i.e. a triaxial structure in the center of the Galaxy,
in order to explain the kinematics of the gas in the center of the Galaxy.

The distance $R_0$ derived here is slightly less that that used in the
model of the disc. However, the small changes in $R_0$ can be compensated
by small changes in the other model parameters, such as the scale
length, so that the predicted counts remain the same. As the model
used already gave a good fit to the disc, we decided not to make {\it ad hoc} 
modifications to account for a smaller $R_0$ since the disc
is not the subject in this paper.

The values for the distance from the Sun to the Galactic centre is very close to the
currently accepted value of just under 8 kpc (see Reid, 1993 for a review).
Recent estimates range from $7.1\pm 1.5$ kpc (Reid et al. 1988) to 
 $8.1\pm 1.1$ kpc (Gwinn, Moran \& Reid 1992).

\section{Bulge stellar density distribution}

The galactocentric distance along the major axis for different isodensity ellipsoids, with the averaged parameters, is

\begin{equation}
t=\sqrt{x^2+K_y^2y^2+K_z^2z^2}
\end{equation}
and the distance along the minor axis is $t/K_z$.
The relationship between the distance $t$ and the density
is given in Table \ref{Tab:dens}.

A power law with exponent $-1.8$ is observed in the centre of the bulge
and also in other galaxies (see review in Sellwood \& Sanders 1988). When  the density function $D(t)$ is fitted  to 
$D(t)=A(t/t_0)^{1.8}exp(-
(t/t_0)^{\gamma })$, with $\gamma $,
$t_0$ and $A$  as free parameters we obtain ($t$ in pc)

\[
D(t)=1.17(t/2180)^{-1.8}\exp(-
(t/2180)^{1.8})\ {\rm stars\ pc}^{-3}
.\]

\begin{table}
\begin{center}
\caption{Relationship between the maximum distance of the ellipsoid
and the bulge star density.}
\begin{tabular}{cccc}
$t$ (pc) & $D$ (pc$^{-3}$) & $t$ (pc)  & $D$ (pc$^{-3}$) \\ \hline
$3020\pm 810$  & $0.1$ & $1620\pm 250$  & $1.1$ \\ 
$2630\pm 650$  & $0.2$ & $1580\pm 240$  & $1.2$\\ 
$2420\pm 590$  & $0.3$ & $1540\pm 240$  & $1.3$ \\ 
$2230\pm 490$  & $0.4$ & $1460\pm 230$  & $1.4$ \\ 
$2120\pm 450$  & $0.5$ & $1420\pm 230$  & $1.5$ \\  
$1990\pm 380$  & $0.6$ & $1390\pm 230$  & $1.6$ \\ 
$1900\pm 350$  & $0.7$ & $1380\pm 240$  & $1.7$ \\ 
$1840\pm 330$  & $0.8$ & $1360\pm 250$  & $1.8$ \\  
$1720\pm 280$  & $0.9$ & $1360\pm 260$  & $1.9$ \\ 
$1670\pm 270$  & $1.0$ & $1320\pm 220$  & $2.0$ \\ 
\label{Tab:dens}
\end{tabular}
\end{center}
\end{table}

\begin{figure}
\begin{center}
\mbox{\epsfig{file=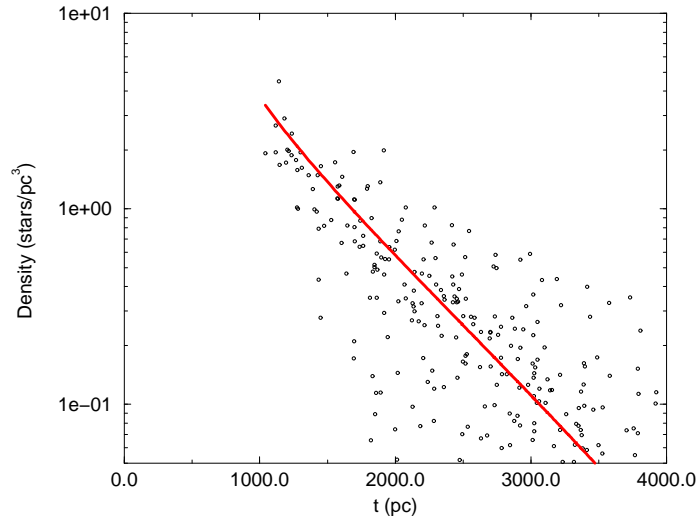,height=8cm}}
\end{center}
\caption{Fit of the density distribution. The solid line is 
the best fit.}
\label{Fig:dens}
\end{figure}

This gives us an estimate of
the fall-off of the density between $1.3$ and $3.0$ kpc from the centre in 
the direction parallel to the major axis or between $0.4$ and $1.0$ kpc in the direction perpendicular to the plane. 
As can be seen in Fig. \ref{Fig:dens}, the dispersion of points
around this law is large, so it is possible to accommodate other functions or
a different parameter set. A different luminosity function would change 
the amplitude of the stellar density. If the normalization for the luminosity function were incorrect then the multiplication factor needed for 
the luminosity function would be used to divide the star density.

\section{Conclusions}

We have found that a)
 the relative abundance of the brightest  sources
in the bulge ($M_K<-8.0$) is much less than in the disc; b) the bulge is triaxial with the major axis nearly along the line of sight to the Galactic centre, the best fit giving a value of
$12$ degrees shifted to positive Galactic longitudes in the plane; and finally
the stellar density drops quickly with distance from the Galactic centre (i.e. the density distribution is sharply peaked).
The $-1.8$ power-law observed at the Galactic centre needs
to be multiplied by an exponential to account for the fast drop in
density in the outer bulge.

The  procedure used here is rather different from that of those authors
who fit the parameters directly to the star counts.
 First, we have inverted the counts.
Then, once we have the luminosity function and  density distribution 
in which the approximate ellipsoidal shape was evident, 
the parameters could be fitted for each isodensity surface. 
Assuming an ellipsoidal bulge with constant parameters 
for all isodensity regions and fitting these parameters to the counts is
less rigorous since there is no {\sl  a priori} evidence for this assumption. 
In fact our method suggest that constant parameters for the 
ellipsoids do not give the best fit for the density $D(\vec{r})$. Instead, a decreasing $K_z(t)$ would provide the best results. 

All these aspects will be developed 
with further details in a future paper, 
where we shall explore the details
and limitations of the inversion and the variation of $K_z$. 

{\bf Acknowledgements:}
We thank the anonymous referee for some helpful comments.







\end{document}